\documentclass[%
aps,
prl,
nofootinbib,
nobibnotes,
amsmath,amssymb,
twocolumn,
]{revtex4-2}
\usepackage{amssymb}
\usepackage{physics}
\usepackage[normalem]{ulem}
\usepackage{graphicx}
\usepackage{amsmath}
\usepackage{amsthm}
\usepackage{amsfonts}
\usepackage{bbm}

\usepackage{xcolor}
\usepackage{hyperref}
\usepackage{overpic}

\usepackage{ifthen}
\usepackage{bbold}
\usepackage{comment}
\usepackage{pdfpages}
\makeatletter
\AtBeginDocument{\let\LS@rot\@undefined}
\makeatother

\newcommand{\newcontent}{}

\newcommand\pN{p_{\mathcal N}^*}

\renewcommand\section[1]{\emph{#1}.---}

\graphicspath{{./imgs}}

\begin{document}

\title{Theory for dissipative time crystals in coupled parametric oscillators}

\author{Stuart Yi-Thomas}
\email[]{snthomas@umd.edu}
\affiliation{Condensed Matter Theory Center and Joint Quantum Institute, Department of Physics, University of Maryland, College Park, Maryland 20742-4111, USA}
\author{Jay D. Sau}
\affiliation{Condensed Matter Theory Center and Joint Quantum Institute, Department of Physics, University of Maryland, College Park, Maryland 20742-4111, USA}

\date{\today}

\begin{abstract}
Discrete time crystals are novel phases of matter that break the discrete time translational symmetry of a periodically driven system. 
In this work, we propose a classical system of weakly-nonlinear parametrically-driven coupled oscillators as a testbed to understand these phases. 
Such a system of parametric oscillators can be used to model period-doubling instabilities of Josephson junction arrays as well as semiconductor lasers. To show that this instability leads to a discrete time crystal we first show that a certain limit of the system is close to 
Langevin dynamics in a symmetry breaking potential. 
We numerically show that this phase exists even in the presence of Ising symmetry breaking using a Glauber dynamics approximation. 
We then use a field theoretic argument to show that these results are 
robust to other approximations including the semiclassical limit when applied to dissipative quantum systems.
\end{abstract}

\keywords{Time Crystal}

\maketitle

The past several years have seen new proposals for possible candidates for phases of matter~\cite{wilczek2012,li2012, else2016}, dubbed ``time crystals'', 
that show spontaneous breaking of time-translation symmetry.
However, later works~\cite{bruno2013,watanabe2015} showed that continuous time-translation symmetry-breaking was not robust, leading to the focus 
on ``discrete time crystals'' (DTCs) in driven Floquet systems which reduce a time-translation symmetry group $\mathbbm Z$ to a subgroup $n\mathbbm Z$%
~\cite{else2016, sacha2015, else2020, khemani2016, moessner2017,  yao2017, machado2023}. 
While an external drive in a closed system 
generically results in an infinite temperature steady state,%
~\cite{dalessio2014, lazarides2014, ponte2015a}
disorder and many-body localization in closed quantum systems%
~\cite{lazarides2015,ponte2015,lazarides2017,khemani2016,else2016,else2020,moessner2017,pizzi2020}
and high frequency drives in classical Hamiltonian systems%
~\cite{hodson2021,abanin2017,else2017,zeng2017,machado2020,natsheh2021, pizzi2021a,pizzi2021b,ye2021,citro2015,yao2020,khasseh2019}
may provide possibilities of dynamical symmetry breaking in {\newcontent a finite} temperature phase.

Another simpler route, which has been the focus of theoretical~\cite{kessler2020,vu2023,passarelli2022,lazarides2020,natsheh2021,nie2023} and 
experimental works~\cite{kessler2021, choi2017, zhang2017, pal2018, rovny2018, ippoliti2021, mi2021, frey2022, kyprianidis2021, beatrez2021, beatrez2023}
in both quantum and classical systems, is to use dissipation to avoid such infinite temperature steady states.  
Such dissipation may lead to DTC order even in classical systems. 
While such behavior might appear to be reminiscent of subharmonic entrainment in classical mechanics~\cite{nie2023,van1927,strogatz2018nonlinear,brown1985harmonic,yu1992fractional}, it is unclear 
to what extent such phenomena are robust to noise as expected in a thermodynamic phase~\cite{hohenberg1977theory}.
In the modern context, classical discrete time crystal (CDTC) behavior appears in coupled arrays of nonlinear pendula 
(e.g.\ the so-called parametrically-driven periodic Frenkel-Kontorova model)~\cite{fk_model, yao2020}, 
{\newcontent though these systems lack a truly infinite autocorrelation time. 
Specifically, work by Yao et al.~\cite{yao2020} utilizes a strong nonlinearity and finds an ``activated'' transition where the autocorrelation time grows with temperature.
Other models, utilizing probabilistic cellular automata (PCAs), do find infinite autocorrelation times~\cite{toom1974nonergodic,gacs2001,gray2001}, even mapping such procedures onto a
Floquet Hamiltonian~\cite{machado2023}, though they have rather complex physical realizations with several-body interactions.}

\begin{figure}[t]
    \centering
    \includegraphics[width=\columnwidth]{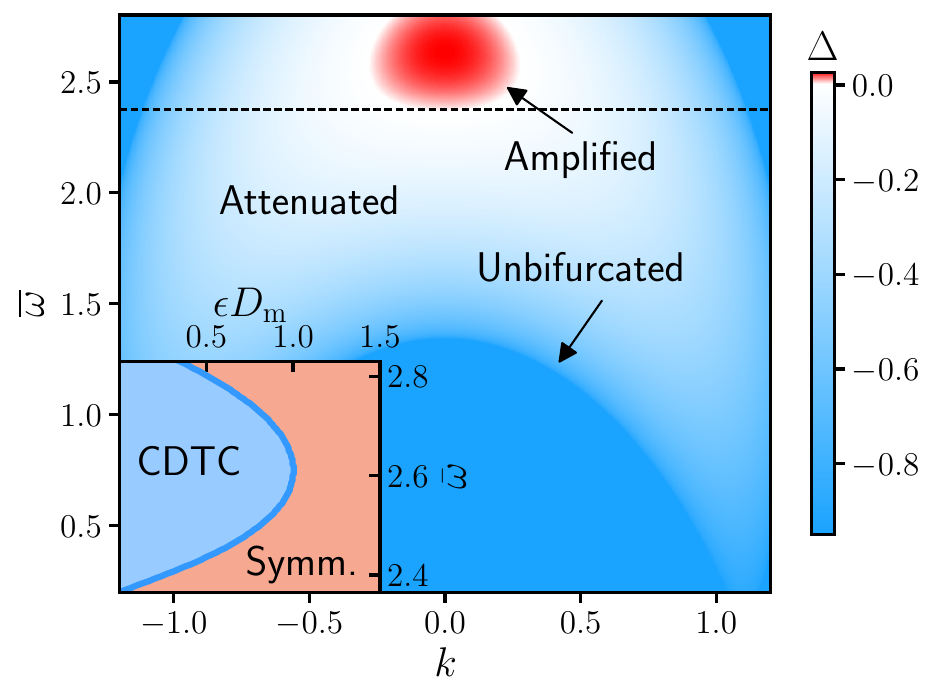}
    \caption{\label{fig:bifurcation} 
        Total gain $\Delta$ of period-doubled state after one period 
        as a function of momentum $k$ and average natural frequency $\overline \omega\equiv(\omega_1+\omega_2)/2$ with damping $\Gamma=0.95$. 
        The gain is given explicitly by $\Delta=\log|\lambda_k| - \Gamma T$, where $\lambda_k$ is the largest Floquet eigenvalue of the oscillator by magnitude.  
        All bifurcated points are in the period-doubled regime, i.e.\ $\mathrm{Re}\, \lambda_k\leq 0$.
        The horizontal dashed line corresponds to the point where the damping is critical, i.e.\ $\Gamma = \Gamma_c=T^{-1}\log|\lambda_{k=0}|$.
    (\emph{inset})~CDTC phase diagram in a 2-dimensional weakly-nonlinear parametric oscillator system based on $\bar\omega$ and the variance of the microscopic noise $\epsilon D_\mathrm{m}$. Parameters from Eq.~\ref{eq:lgw-hamiltonian} are compared with numerical results from \cite{schaich2009}. %
    $J=10$, $|\omega_1-\omega_2|/2=0.664\pi$, %
    and $g=\nu=T=1$.
     }   
\end{figure}

{\newcontent In this work we propose an absolutely stable CDTC in an array of parametrically-driven perturbatively-nonlinear dissipative oscillators and use this construction to study a CDTC phase. Using a weak nonlinearity, this system exhibits an infinite autocorrelation time in experimentally-tractable system.}
Similar parametric oscillator systems have been shown to realize period-doubling bifurcations in semiconductor laser systems~\cite{simpson1995}, {\newcontent nanoelectromechanical systems~\cite{sarkar2023} } and Josephson junction arrays~\cite{vijaybifurcation}, all of which can operate in 
both dissipative quantum and classical limits.
As shown in Fig.~\ref{fig:bifurcation},
for an appropriate choice of the average natural frequency, the array of oscillators shows a period-doubled amplifying phase over a small range of wavevectors $k$.  

We show that, for an appropriate choice of weak nonlinearity {\newcontent(relative to previous work~\cite{yao2017,yao2020,machado2020})} and noise in the classical limit, the above system shows a symmetry-breaking transition that can be described by 
Ginzburg-Landau theory. 
The CDTC order features a time scale that scales exponentially in system size, similar to equilibrium symmetry breaking~\cite{goldenfeld2018, griffiths1966relaxation,tomita1992statistical}.
{\newcontent We use a Monte Carlo simulation to demonstrate that this phase is robust to perturbations which break any preexisting symmetries of the system, demonstrating that discrete time-translation symmetry can be broken without another symmetry. 
We then generalize the field-theory description of the time-crystal phase~\cite{else2017} to the finite temperature dissipative phases obtained here.}

\section{Period-doubling in coupled parametric amplifiers}\label{sec:amplifiers}
We start by considering a coupled array of parametric oscillators with translational invariance in $\mathcal D$ spatial dimensions. We choose to model such an array by a time-dependent Hamiltonian
\begin{align}
    \label{eq:hamiltonian}
H(t) &= \sum_j \frac{1}{2} \dot \phi_j^2 + \frac{1}{2} \omega^2(t) \phi_j^2 + \frac{J}{2} \sum_{\hat\mu}^{\mathcal D} \left( \phi_j-\phi_{j+\hat\mu} \right)^2.
\end{align}
where $\omega(t+T)=\omega(t)$ is periodic in time and $\hat\mu$ iterates over each lattice vector.
An individual periodically modulated oscillator can exhibit a parametric resonance~\cite{parametricamplification,floquetanalysis} featuring ``amplified'' and ``squeezed'' modes. As an illustration, we consider a piecewise-constant time dependence of the driving frequency
$\omega(t) = \omega_1 \Theta(t/T - 1/2) + \omega_2 \Theta(-(t/T - 1/2)) $,
which is periodic in time with period $T$.
Due to translation invariance, we can write the dynamics in terms of Bloch momentum modes $\psi_k=(\phi_k, \dot \phi_k)^T$. The Floquet time-evolution over period $[0,T]$ is given by a total transfer matrix $M_k$ which equals the product of two $2\times 2$ transfer matrices $M_k = M_{k,2}\, M_{k,1}$, each given by %
time-evolution of a harmonic oscillator with frequency $\omega_{j=1,2}$ and time $T/2$. Explicit forms are found in the Supplementary Material (SM).
For an appropriate choice of model parameters, the transfer matrix $M_k$ features amplified and squeezed modes, denoted $\psi^\mathrm{A}_k$ and $\psi^\mathrm{S}_k$, with real eigenvalues $\lambda_k$ and $\lambda^{-1}_k$. Without loss of generality, we assume $|\lambda_k| \geq 1$.

When the eigenvalue $\lambda_{k}$ is less than $-1$ for some $k$, the parametric amplification can lead to an instability towards a period-doubled state where
the parametrically amplified normal mode $\psi^\mathrm{A}_k(t)$, as well as the corresponding displacement $\phi_k^\mathrm{A}(t)$, switches sign between periods. Choosing the parameter $J$ such that $\lambda_k$ has a maximum at $k=0$ leads to a spatially-uniform period-doubled extended state being dominant.
However, the system is unstable to perturbations since $|\lambda_k|>1$.
We remedy this by adding a damping to the equations of motion (a force proportional to $-\dot\phi$) such that the eigenvalues become $\lambda_k\rightarrow e^{-\Gamma T} \lambda_k$. 
{\newcontent Such a damping is naturally realized in Josephson junctions through a normal metal resistor or coupling to a transmission line, or in semiconductor laser systems from the finite quality factor of the cavity.}
At a critical damping $\Gamma_c = T^{-1}\log \lambda_{k=0}$, only the $k=0$ period-doubled state persists.
The steady state dynamics at this fine-tuned point features displacements $\phi_j(t)$ that are uniform in space  and satisfy the period-doubling condition $\phi_j(T)=\phi_0(T)=-\phi_j(0)$ for all lattice sites $j$.
Though this system shows period-doubled dynamics without any instability, it does not recover from applied perturbations and therefore does not satisfy the robustness criteria of a CDTC; the addition of weak noise and anharmonicity can stabilize this state into a true CDTC.

\section{Perturbative nonlinearity and noise}
We perturb away from the fine-tuned point by setting the damping to $\Gamma=\Gamma_c-\epsilon (\nu/2T)$ where $\nu$ is a constant factor which we add for later convenience and $\epsilon\ll1$. This change causes a small region around $k=0$ to be unstable, shown in Fig.~\ref{fig:bifurcation}. Motivated by model A dynamics of a symmetry breaking system~\cite{hohenberg1977theory}, we add a small nonlinear force 
\begin{align}
  \label{eq:nonlinforce}
f_k(t) &= \epsilon \sum_j e^{-ikj} \left[ -g \phi_j(t)^3 + \frac \nu {2T} \dot \phi_j(t) + \eta_j(t) \right].
\end{align}
which is proportional to $\epsilon$ 
to stabilize the amplification in this region.
In this expression, the terms represent a restoring nonlinear force, the reduction in damping, and uncorrelated Gaussian noise with variance $\langle \eta_i (t) \eta_j(t') \rangle = 2D_\mathrm{m} \delta_{ij}\delta(t-t')$ respectively.
{\newcontent Since $\epsilon$ is small, the oscillatory micromotion is approximately exact while the slow dynamics is well approximated by discrete samples at times $nT$. This is analogous to the separation of dynamics in quantum Floquet systems~\cite{eckardt2015}.}
Here $g$ sets the positions of the fixed points generated by the nonlinear restoring force.
{\newcontent While the nonlinearity is crucial for ensuring stability, the noise is also important to avoid the complexities of nonlinear many-body Hamiltonian dynamics, averting singularities by broadening transition probabilities~\cite{van1927,strogatz2018nonlinear}

Applying the force $f_k(t)$, the %
evolution for $\psi_k$ over a time-step $2 nT$ is written as
 \begin{align}
\psi_k (2nT)&=\left(1+\nu\epsilon/2\right)^{2n} M_k^{2n} \psi_k(0) \notag \\
 &+ \epsilon \int_0^{2nT} d\tau\, G_k(2nT,\tau)e_2 F_k[\{\psi_k\}, \tau] \notag \\
 &+ \epsilon \int_0^{2nT} d\tau\, G_k(2nT,\tau) e_2 \eta_k(\tau)\label{eq:evolve} \\
 F_{k}[\{\psi_k\},t] &\equiv -g\int dk_{1,2}\, \phi_{k_1}(t) \phi_{k_2}(t) \phi_{k-k_1-k_2}(t). \notag
\end{align}
Here, the Greens function $G_k$ generalizes the transfer matrix to continuous times ($G_k(nT,0)=M_k^n$) and describes the fast dynamics of the parametric oscillators. The unit vectors $e_2=(0,1)^T$ pick out the momentum component of the Greens function matrix.

Choosing $n$ to be  sufficiently large so that the squeezed eigenvalue of $M_k^{2n}$ becomes comparable to $\epsilon$ for all momenta
we observe that the total amplitude of the squeezed mode in $\psi_{k}(2nT)$ must be on the order of $\epsilon$.
The squeezed mode's effect on the amplified mode via the nonlinearity is then proportional to $\epsilon^2$ and we can safely ignore its contribution to understand the steady state dynamics of the amplified mode, taking $\psi_k \simeq \psi^\mathrm{A}_k$ and $\phi_k \simeq \phi^\mathrm{A}_k$.
Using this assumption, we can write the steady-state equation of motion over two periods as
\begin{multline}
  \label{eq:A_mode}
\phi_k(2T) =\left(1+\nu\epsilon\right)  \lambda_k^2 \phi_k(0) + \epsilon \zeta_k \\
         - \epsilon g \int dk_{1,2}\, \xi_{k,k_1,k_2}\, \phi_{k_1}(0) \phi_{k_2}(0) \phi_{k-k_1-k_2}(0)
\end{multline}
to first order in $\epsilon$.
The coefficient $\xi_{k,k_1,k_2}$ and the effective noise $\zeta_k$ encode the nonlinear effects and the stochastic process over two driving periods.

Scaling the momenta $k\rightarrow k\sqrt{\epsilon}$, which we can safely perform in the thermodynamic limit,  allows us to expand the momentum-dependent amplification rate as
\begin{equation}
(1+\nu\epsilon)\lambda_k^2=(1 + \nu\epsilon) - \epsilon \gamma k^2 + O(\epsilon^2).
\end{equation}
Furthermore, the coefficient for the nonlinear term and the noise strength both become $k$-independent up to leading order in $\epsilon$, allowing us to replace $\xi_{k,k_1,k_2}$ with $\xi \equiv \xi_{0,0,0}$ and assert that $\zeta_k$ has a constant variance. 
{\newcontent
With these assumptions, focusing on the displacement component of $\psi_k^\mathrm{A}$, we can approximate the discrete map in Eq.~\ref{eq:A_mode} as continuous dynamics by rescaling $T\rightarrow T\epsilon$. This yields a real-space Langevin equation 
}
\begin{align}
  \label{eq:langevin}
 \dot \phi_i = - {\newcontent\frac 1 {2T}} \left( \frac {\delta H} {\delta \phi_i} + \zeta_i \right)
\end{align}
with the Ginzburg-Landau (GL) Hamiltonian
\begin{align}
  \label{eq:lgw-hamiltonian}
  H[\{\phi_{i}\}] = \sum_{i} -\frac \nu {2} \phi_i^2  + \frac {\gamma} {4} (\nabla \phi_i)^2 + \frac { g \xi } {4} \phi_i^4.
\end{align}
In this form, Eq.~\ref{eq:langevin} is also known as a time-dependent Ginzburg-Landau equation. 
{\newcontent The variance of noise $\zeta_i(t)$ is $4T\epsilon D$, which includes a factor of $2T\epsilon$ from the conversion to a function of continuous time.
(Explicit formulae for $D$ and $\xi$ are given in the SM.)}

The Langevin equation relaxes the system towards the Boltzmann distribution associated with the GL Hamiltonian \cite{goldenfeld2018}
\begin{equation}
  p\big[\{\phi_{k}\}\big]\sim e^{-H[\{\phi_{k}\}]/\epsilon D}.\label{eq:Boltzmann}
\end{equation}
This steady state distribution, for systems with dimension two or higher, shows spontaneous symmetry breaking of the $\phi\rightarrow -\phi$ Ising symmetry when the parameter $\nu$ in Eq.~\ref{eq:lgw-hamiltonian} is greater than a critical value $\nu_c$. 
In this case, the dynamics on the even time steps $2nT$ breaks ergodicity~\cite{goldenfeld2018} into configurations 
\begin{equation}
    \langle \phi_i(2 nT)\rangle =\textrm{constant} \label{eq:order},
\end{equation}
breaking the Ising symmetry of the Hamiltonian $H$. (At odd time steps, this constant has the oppostive sign since the eigenvalue $\lambda_k$ is less than one.)
Using known data on the GL phase transition on a 2D lattice~\cite{schaich2009}, we calculate the phase boundary, delineated by  $\nu=\nu_\mathrm{c}$, 
in terms of microscopic parameters, shown in the inset of Fig.~\ref{fig:bifurcation}.
This transition can be extended to one dimension 
by introducing long-range interactions $O(1/r^\alpha)$  for $1<\alpha<2$~\cite{dyson1969existence}.
We therefore expect a CDTC phase transition at a critical noise strength when $1<\alpha<2$, which is consistent with the result cellular automata result from~\onlinecite{pizzi2021}.
{
\newcontent

\section{Robustness}
The Ising symmetry $\phi\mapsto -\phi$ played a crucial role in the GL treatment of symmetry breaking described in the previous section. 
However as a matter of principle, the time-crystal phase should be robust to perturbations that break this symmetry~\cite{keyserlingk2016}.
To demonstrate that the time-crystal phase is indeed robust, we describe the time crystal phase in this system without any reference to an Ising symmetry through an order parameter
 \begin{align}
     \label{eq:time-crystal-order-parameter}
     O=\bar{\phi}(2nT) - \bar{\phi}((2n-1)T)
 \end{align}
 whose expectation value is a measure of time translation symmetry breaking and  $\bar{\phi}=L^{-d}\sum_i \phi_i$ is the average magnetization over the system. 
 This time-crystal order parameter is non-zero in the Ising symmetric 
 GL system described by Eq.~\ref{eq:order} together with the period doubling condition $\phi_i(2 n T)\simeq -\phi_i((2n-1)T)$.

We verify this robustness using a Markov chain Monte Carlo simulation of overdamped dynamics 
by replacing the Langevin dynamics in Eq.~\ref{eq:langevin} by Glauber dynamics~\cite{glauber1963} on the GL Hamiltonian in Eq.~\ref{eq:lgw-hamiltonian} 
together with a symmetry breaking term $\sum_i d\phi_i^3$. 
This implements a discrete dynamics similar to Eq.~\ref{eq:A_mode} rather than the continuous time limit 
in Eq.~\ref{eq:langevin}.
 A similar approach was taken in Ref.~\onlinecite{gambetta2019}.
While the Glauber dynamics, in general, differs from the oscillator evolution (i.e. Eq.~\ref{eq:evolve}), it reproduces the GL stationary distribution obtained from 
Eq.~\ref{eq:A_mode} in the limit of small $\epsilon$ so that the dynamics is easy to calibrate.
We then model the period doubling property of each oscillator as a global flip $\phi \mapsto -\phi$ at each time step $T$.
We measure the time crystal order parameter $O$ on a 2D lattice and find a phase transition that persists for finite $d$ in Fig.~\ref{fig:monte-carlo}, though with a different critical point. Due to the asymmetry, the time-averaged magnetization $\langle \bar\phi \rangle$ deviates from the symmetric point at $\bar\phi=0$, indicating that the Ising symmetry is truly broken. 
This behavior indicates that the periodic drive creates an emergent Ising symmetry which is analogous to that of the $\pi$-spin glass system in the quantum case~\cite{khemani2017,keyserlingk2016}.
The phase transition remains robust up until $d\approx 1$, where the system no longer features time-crystalline behavior. More details are found in the Supplementary Material.

\begin{figure}[t]
\centering
\begin{overpic}[width=\columnwidth]{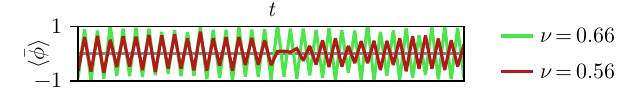}
\put(0,0.7){(a)}
\end{overpic}
\label{elephant}
\begin{overpic}[width=\columnwidth]{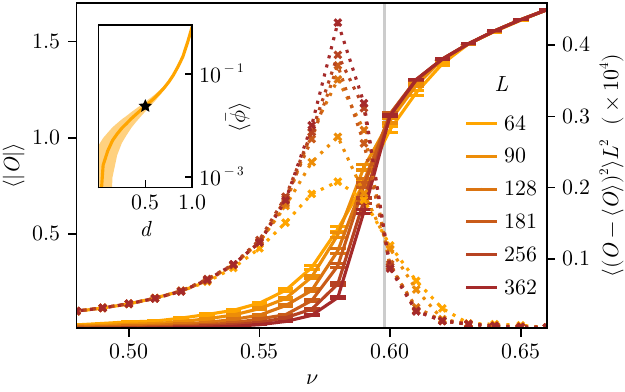}
\put(0,0.7){(b)}
\end{overpic}
    \caption{\label{fig:monte-carlo} \newcontent
     (a) Demonstration of phase-slip in the symmetric phase (red) compared with broken phase (green); (b) Order parameter $O$ (Eq.~\ref{eq:time-crystal-order-parameter}, solid lines) and susceptibility (dotted lines) in the 2D asymmetric $\phi^4$ model demonstrating a second-order phase transition in the quadratic coefficient at $\nu_\mathrm{c}\approx 0.598$ (grey line) with an asymmetry of $d=0.5$, compared to $\nu_\mathrm{c}=0.72112$ for the symmetric model \cite{schaich2009}.
    The inset shows time-averaged magnetization $\langle\bar\phi\rangle$ as a function of asymmetry strength, with a black star at $d=0.5$ denoting the value for the main plot. $\gamma=4$, $g\xi=0.5$, $\epsilon D=2$. 
     }   
\end{figure}

To further understand these results, we formally define the coarse-grained Markov chain dynamics at the time steps $t=n T$ as %
\begin{align}
    &\psi_{i}((n+1)T)=Q_{i}\big[\{\psi_j(nT)\}\big]\label{eq:PCA}
\end{align}
where $Q_i$ is a random functional of the displacement and momentum configurations $\{\psi_j\}$. 
Since $Q_i$ is local, the process is equivalent to a local Markov process or a PCA.
This 
dynamical map generates a Markov process with a transition probability
\begin{align}
    \mathcal L\big[\{\psi'_i\}|\{\psi_j\}\big] = \Big\langle \prod_i \delta\left( \psi'_i - Q_i\big[\{\psi_j\}\big] \right) \Big\rangle.
    \label{eq:transition-probability}
\end{align}
that evolves the probability distribution between times $nT$ and $(n+1)T$. Here, $\delta$ is the Dirac delta function.
For simplicity, we will represent $\mathcal L$ as a linear operator on the space of probability distribution vectors $p[\{\psi_k\}]$ through the relation 
\begin{equation}
\left(\mathcal Lp\right)\big[\{\psi_k\}\big]\equiv \int D\psi'_{k'}\, L\big[\{\psi_k\}|\{\psi'_{k'}\}\big]\,p\big[\{\psi'_{k'}\}\big].\label{eq:Lp}
\end{equation}
The above equation can also be viewed as an evolution equation for a quantum system with a Markovian dissipation (e.g.\ via a Lindblad master equation~\cite{passarelli2022}) if the 
probability $p\big[\{\psi_{k}\}\big]\rightarrow p\big[\{\psi^*_{-k},\psi_{k}\}\big]$ is a Wigner representation of the quantum density matrix. Note that in the quantum case, it is more convenient to change the fundamental variables to  
$\psi_k\simeq\phi_k+i\dot{\phi}_k$ which would represent the oscillator position and momenta in a coherent state representation. The linear operator $\mathcal L$ 
is the time-evolution of the density matrix over $T$, which is an exponential of the Lindblad superoperator. 
The classical equations of motion Eq.~\ref{eq:evolve} can be obtained from the truncated Wigner approximation~\cite{polkovnikov2003quantum} 
of the quantum evolution in the large occupation/weak interaction limit of the oscillator modes. 
We will allow $p[\{\psi_k\}]$ to be a density matrix in the rest of the work.

Since the transition probabilities depend only on the instantaneous state of the system, the dynamics form a Markov chain that is finite\footnote{Though we are ultimately interested in the thermodynamic limit, any given system size is finite in the mathematical sense.} and ergodic and therefore admits a stationary probability distribution $p^*[\{\psi_i\}]$ (the asterisk here denotes stationarity, not complex conjugation). 
In the quantum case, $p^*$ is the stationary solution 
of the quantum master equation.
Taking the logarithm of this distribution gives an effective Hamiltonian
\begin{equation}
H[\{\psi_i\}] \propto -\log p^*[\{\psi_i\}]\label{eq:Heff}.
\end{equation}
This is similar to the effective Hamiltonian of pre-thermal quantum systems~\cite{else2017}.
To understand magnetic phases, it is more insightful to consider the probability distribution of the 
magnetization $\bar{\phi}$ written as: 
\begin{equation}
P(M) = \left\langle \delta\left(M-\bar{\phi}\right) \right\rangle_{p^*}\label{eq:PM}.
\end{equation}

   In a system without continuous symmetries, for short range correlated phases, the central limit theorem suppresses 
   the variance of the mean magnetization. 
   Systems that are not critical are typically short range correlated except when the system 
   breaks symmetry spontaneously and the system separates into phases with different mean magnetizations~\cite{goldenfeld2018}.
   We thus expect $P(M)$ to be sharply peaked in the thermodynamic limit.
   Specifically, $P(M)$ approaches a weighted sum of Dirac delta functions in the thermodynamic limit. 
   This is analogous to coexistence in equilibrium spontaneous symmetry breaking.
   
   We now pick one delta peak at $M=M_1$ and choose a neighborhood $\mathcal N$ in the space of magnetizations such that $M_1$ is the only peak in $\mathcal N$.
   We then define the conditional probability distribution
   \begin{equation}
   \pN\left[\{\psi_i\}\right]\equiv p^*\big[\{\psi_i\} \,|\, \bar\phi \in \mathcal N\big]\label{eq:pN}
   \end{equation}
   which does not exhibit coexistence or long range order. 
   The corresponding density matrix in the quantum case, which would be defined by applying a projector $\mathcal{P}_{\bar\phi \in \mathcal N}$,
   is analogous to the symmetry breaking density matrix in the prethermal case~\cite{else2017}.
   The magnetization after one time step is then
    \begin{equation}
    \label{eq:def-M2}
    M_2 = \langle \phi_i \rangle_{\mathcal L \pN}.
    \end{equation}
    If $M_2\neq M_1$, then this map defines a distinct delta peak in $P(M)$ around $M_2$.
   One can check that in this case, if $\pN$ is stationary under the Markov chain $\mathcal L^2$, then $\mathcal L$ defines a period-2 discrete time crystal. 
   Due to the peaked structure of $p^*$, we can relax these conditions to
    \begin{align}
    \langle \bar\phi \rangle_{\mathcal L \pN} &\notin \mathcal N, && \langle \bar\phi \rangle_{\mathcal L^2 \pN} \in \mathcal N. \label{eq:conds}
    \end{align}
    
    While it is clear that these conditions lead to DTC behavior in infinite systems, we claim that satisfactory systems for any size $N$ must have a diverging autocorrelation time as $N\rightarrow\infty$. 
    In a large but finite-sized system, we can soften the definition of $\mathcal N$ to including only one maximum of $P(M)$. 
   For example, $\mathcal{N}$ can be chosen to be the region around $M_1$ over which $P(M)$ is concave, which we expect roughly maximizes the likelihood that the conditions correctly identity CDTC behavior.
    Since the magnetization is intensive, the positions of the peaks are independent of system size and $\mathcal N$ encompasses only one peak as $N\rightarrow \infty$, satisfying the aforementioned condition. 
    We note that this argument may not extend to period-$n$ symmetry breaking for $n>2$ \cite{bennett1990}.
    
    In a finite-sized system, a phase slip in the CDTC order $O$ occurs if the magnetization at any given time step falls inside the neighborhood $\mathcal N$ after one time-step or outside after two time-steps. %
    Assuming Eq.~\ref{eq:conds}, such a phase slip may have a nonzero probability due to the variance of $\bar\phi$. The probability of a phase slip is related to the variance of $\bar\phi$ under $\mathcal L^n \pN$ for $n=0$, $1$ and $2$, which decays exponentially in $L^{d-1}$ (as opposed to $L^d$, due to domain walls).
}

{\newcontent Incidentally, the temperature of the classical system considered here needs to be low (i.e.\ $o(\epsilon)$) since the microscopic noise power scales as $\epsilon$ (since $\epsilon D_\mathrm{m}\sim o(1)$ in Eq.~\ref{eq:Boltzmann}) while the microscopic damping scales as unity. These considerations should be relevant to evidence for dissipative quantum time crystals in recent experiments~\cite{kessler2021}. 
}

\section{Discussion}
We have considered the possibility of a CDTC state in a coupled array of period-doubling parametric amplifiers and explained how our results can be robust in a more general context of PCAs. 
Since parametric amplifiers occur in many driven, weakly dissipative nonlinear systems such as Josephson junctions~\cite{vijaybifurcation, yaakobi2013parametric, russomanno2023}, nonlinear optics~\cite{galiffi2022photonics} and nonlinear electronic circuit elements~\cite{ranganathan2003discrete}, we are optimistic that this analysis motivates experimental studies of CDTC symmetry breaking in more systems.

Using a course-grained mapping to Glauber dynamics that was solved numerically as well as a field-theoretic~\cite{else2017} argument, we have 
shown that the CDTC phase obtained is absolutely stable~\cite{machado2020} and robust to Ising symmetry breaking and other perturbations. 
{\newcontent The field-theoretic argument implies conditions for identifying a CDTC that apply to translation-invariant systems without continuous symmetries and with a spatially self-averaging order-parameter. 
Additionally,
the classical system discussed can be viewed as the semiclassical limit of the corresponding quantum system in the so-called truncated Wigner approximation~\cite{polkovnikov2003quantum}. The robustness argument using the field-theoretical formalism extends to quantum corrections. The central difference between our analysis and the one in closed quantum systems~\cite{else2017},  where the CDTC phase is prethermal, appears to be the presence of dissipation instead of disorder. The presence of disorder can complicate the identification of an order parameter~\cite{else2017}.
A better understanding of CDTC systems with disorder and dissipation is an interesting future direction.
}

\begin{acknowledgments}%
We thank Francisco Machado for alerting us to the emergent Ising symmetry in time crystals. We also thank Sankar Das Sarma, DinhDuy Vu, 
Tom Iadecola and Paul Goldbart for valuable comments on the work.
S.Y.T.\@ thanks the Joint Quantum Institute at the University of Maryland for support through a JQI fellowship.
J.S.\@ acknowledges support from the Joint Quantum Institute as well as valuable discussions 
hosted at the Aspen Center for Physics, which is supported by National Science Foundation grant PHY-2210452.
This work is also supported by the Laboratory for Physical Sciences through its continuous support of the Condensed Matter Theory Center at the University of Maryland.
\end{acknowledgments}%

\bibliography{bib.bib}

\newpage
\includepdf[pages={1,{},{},2,{},3}]{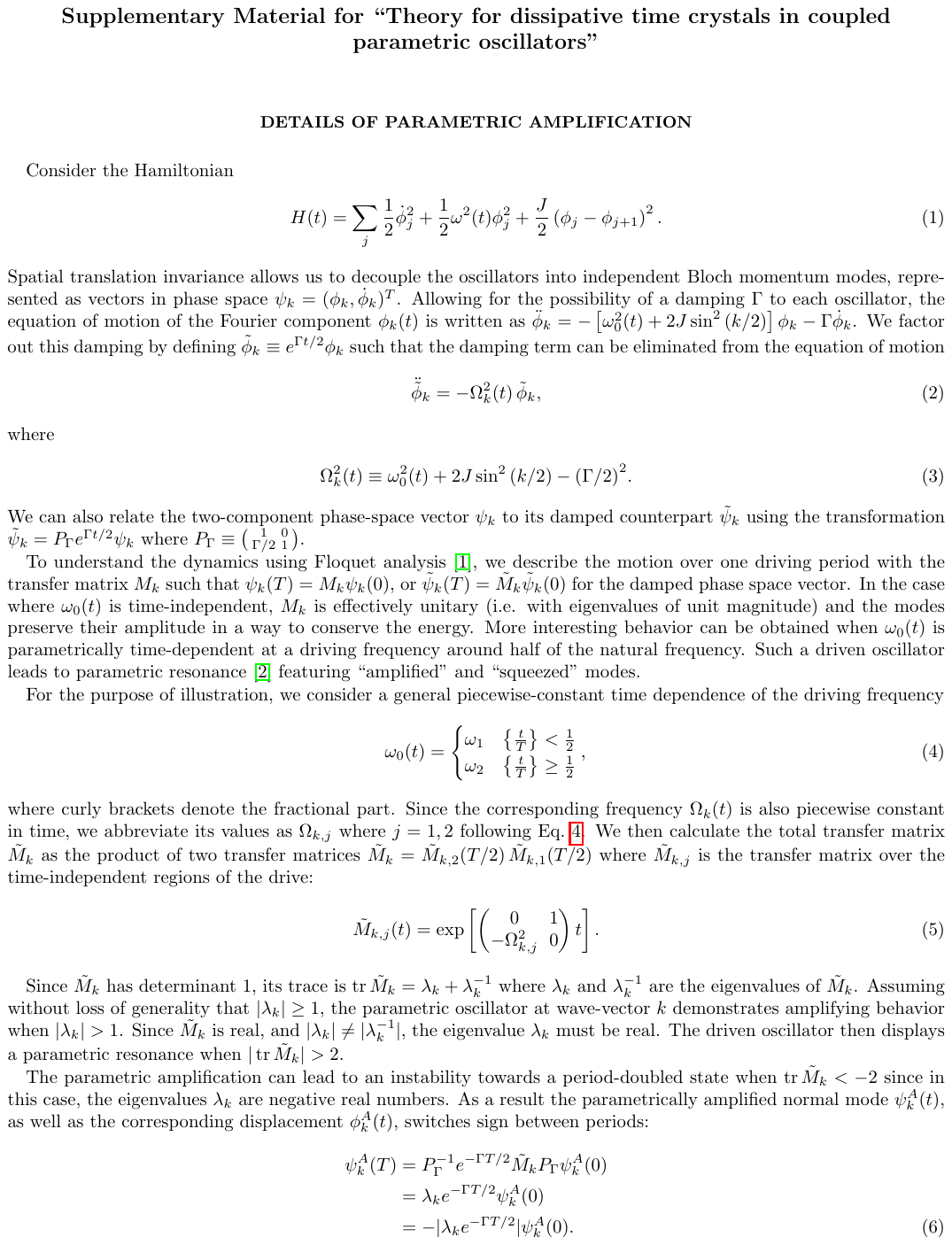}

\end{document}